\documentclass[12pt]{iopart}
\usepackage{graphicx}
\usepackage{iopams}
\usepackage{color}

\def\w{\omega}

\def\vp{\varphi}
\def\e{\varepsilon}

\newcommand{\eqref}[1]{(\ref{#1})}

\begin{document}
\title[Effective phase connectivity from observations]
{Reconstructing effective phase connectivity of oscillator networks from
observations}
\author{Bj\"orn Kralemann$^1$, Arkady Pikovsky$^{2,3}$, and Michael Rosenblum$^2$}
\address{$^1$ Institut f\"ur P\"adagogik, Christian-Albrechts-Universit\"at zu
Kiel,
Olshausenstr. 75, 24118 Kiel, Germany}
\address{$^2$ Institute of Physics and Astronomy, 
University of Potsdam, Karl-Liebknecht-Str. 24/25, 14476 Potsdam-Golm, Germany}
\address{$^3$ Department of Control Theory, Nizhni Novgorod State University,
Gagarin Av. 23, 606950, Nizhni Novgorod, Russia}
\begin{abstract}
We present a novel approach for recovery of the directional connectivity of a small 
oscillator network by means of the phase dynamics reconstruction from 
multivariate time series data. 
The main idea is to use a triplet analysis instead of the traditional pairwise
one. 
Our technique reveals an effective phase connectivity which is generally 
not equivalent to a structural one.
We demonstrate that by comparing the coupling functions from all possible 
triplets of oscillators, we are able to achieve in the reconstruction
a good separation between existing and non-existing 
connections, and thus reliably reproduce the network structure.
\end{abstract}

\section{Introduction}
Understanding and mapping of the brain connectivity is one of the most
challenging problems of neuroscience
\cite{10.1371/journal.pcbi.1000334,Bullmore-Sporns-09,10.3389/fninf.2012.00014,
Boly_et_al-12,Pastrana-13,Sporns-13}. 
The information provided by the matrix of inter\-connections of the human brain,
or ``connectome'' 
\cite{10.1371/journal.pcbi.0010042}, is essential for both basic and applied
neurobiological research.
It is believed that numerous severe disorders like Parkinson's and Huntington's
diseases, autism, 
schizophrenia, and dyslexia are related to disorders in brain connectivity 
(see \cite{10.1371/journal.pcbi.1000334} and references therein). 
A particular important problem is to reveal connectivity, and especially 
\textit{directional connectivity},
from multivariate data, e.g. from multichannel electroencephalography (EEG) or 
magnetoencephalography (MEG) measurements \cite{Lehnertz-11}. 
Another, but related, 
problem is to find out whether two correlated nodes are connected directly or indirectly. 
In a more general context, the problem of revealing connectivity can be formulated
as a general data analysis problem for oscillator networks: how to recognize
the network structure, i.e. the directions and the strengths of the couplings,
from the observation. 
In particular,
similar tasks arise also in the analysis of other physiological systems, 
e.g. in quantification of cardio-respiratory interaction 
\cite{Mrowka_et_al-03,Musizza_et_al-07,Kralemann_et_al-13} 
and in a number of other fields, e.g. in climate physics 
\cite{Sharma_et_al-12,Wang_et_al-12,PhysRevLett.108.258701} and ecology \cite{Sugihara26102012}.

Various techniques, mostly based on different measures of correlation 
\cite{Skudlarski-08}, synchronization
\cite{Tass_et_al-98,Rodriguez_et_al-99,Axmacher2008,Bartsch26062012}, 
and mutual or transfer information 
\cite{Schreiber-00,PhysRevE.67.055201,PhysRevE.75.056211,PhysRevLett.99.204101,PhysRevE.76.036211,%
PhysRevLett.100.084101,PhysRevE.77.026214,PhysRevLett.100.158101,PhysRevLett.103.238701,%
PhysRevE.83.011919,PhysRevE.83.051122,PhysRevE.83.051112,%
Chicharro-Andrzejak-Ledberg-11,Battaglia-Witt-Wolf-Geisel-12,PhysRevE.86.061121,Kugiumtzis-13} 
have been exploited to tackle the connectivity problem.
A separate group of methods relies on the coupled oscillator theory; these
techniques assume that the nodes 
of the network are active, self-sustained, oscillators and reveal directional
connectivity by reconstructing the phase 
dynamics 
\cite{Rosenblum-Pikovsky-01,Kralemann_et_al-07,Kralemann_et_al-08,Axmacher2008,
Cadieu2010,dcm_phase,%
Kralemann-Pikovsky-Rosenblum-11,PhysRevLett.109.024101,PhysRevE.73.031915} via analysis of 
instantaneous phases and instantaneous frequencies.
In this paper we further develop the phase dynamics  approach, 
by substituting the pairwise analysis of
a network by an analysis 
of triplets of nodes. We show that this essentially improves
the reliability of the method. Although motivated by the problems of neuroscience, the
technique can be exploited to 
recover the structure of networks of various nature.

\section{Structural, functional, and effective phase connectivity}
\label{sec2}
\begin{figure}[ht!]
\centerline{\includegraphics[width=0.8\textwidth]{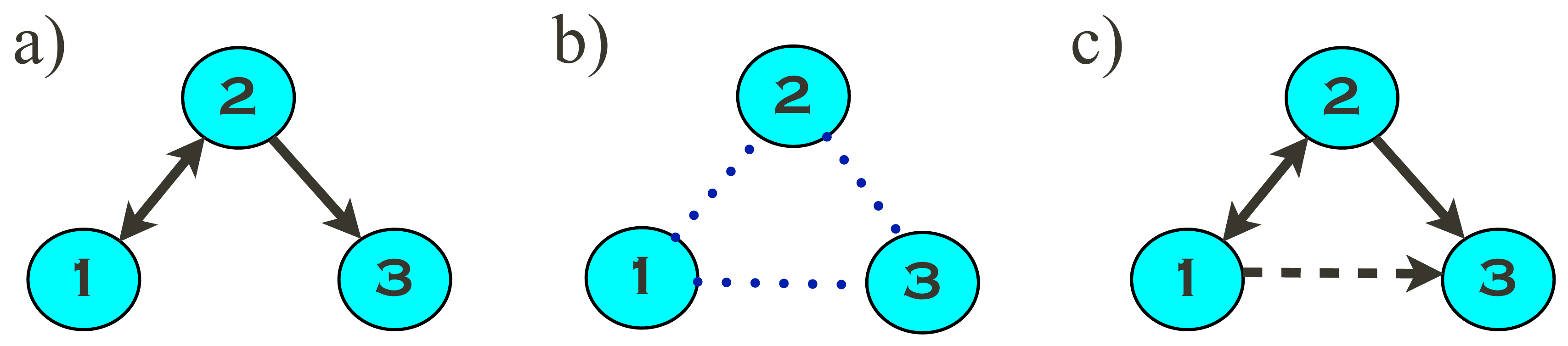}}
\caption{Illustration of the notions of connectivity. 
(a) Nodes $1,2$ and $2,3$ are physically (structurally) connected. 
This connection can be bi- or unidirectional, as shown by arrows.
(b) Structural connection causes correlated (or synchronous) dynamics, i.e. functional 
connection (dotted lines).
Since measures of correlation and synchronization are symmetric, functional connectivity is
not  directional.
Notice that although nodes $1,3$ are not physically connected, they may be correlated due 
to the common drive from the node $2$ and indirect action of node $1$, which yields an additional 
link if compared with (a).
(c) If there is a directional structural link $k\to l$, then phase dynamics of node $l$ depends on 
the node $k$, i.e. the same link exists in the effective phase connectivity map.
However, additional links may appear. So, for this example, phase dynamics of node $3$ may depend 
on that of node $1$, i.e.  these nodes may be also effectively connected (dotted line). 
Notice that  phase connectivity is also directional.
So, e.g., the $1\to 3$ effective phase connection is possible in this
configuration, while the $3\to 1$ connection is not. 
}
\label{connillustr}
\end{figure}

Before proceeding with  the presentation of the technique we discuss different
notions of connectivity. 
In neuroscience one typically differentiates between anatomical connections, 
functional connections, corresponding to 
correlations between pairs of nodes which may be not connected anatomically, and
effective connections, 
representing direct or indirect causal influences of one node on the other
\cite{Friston-11,Rubinov-Sporns-10}. 
We illustrate these important notions by the following example. 
Suppose we have three self-sustained oscillators, coupled as shown in
Fig.~\ref{connillustr}a.
Here the arrows indicate physically existing, uni- or bidirectional,
connections. 
For physical systems these connections are implemented by means of resistors,
optical fibers, etc. 
For biological systems they correspond to anatomical connections, e.g. via
synapses
or via white matter tracts between pairs of brain regions
\cite{Rubinov-Sporns-10}. 
Mathematically, this means that, e.g., for the second node we can write
\begin{equation*}
\mathbf{\dot x}_2=\mathbf{G_2(x_2)}+\e\mathbf{H}_2(\mathbf{x}_2,\mathbf{x}_1)\;,
\end{equation*}
where the vectors $\mathbf{x}_k$ are state variables of the $k$th
oscillator, 
functions $\mathbf{G}$ and $\mathbf{H}$ describe the autonomous dynamics and the
coupling, respectively;
the coupling parameter $\e$ explicitly quantifies the strength of interaction.
Since the dynamics of the second oscillator depends on the state of the first
one, we say that there is 
a \textit{structural connection}  $1\to 2$. Obviously, the structural
connectivity is directional.

Next, consider two physically uncoupled nodes $1$ and $3$. 
Since node $1$ acts indirectly on the node $3$ via the node $2$ and both nodes $1,3$ have a common drive,  
they can synchronize or, more generally, can oscillate with a certain degree of correlation. 
A measure of correlation or synchrony would reflect the non-directional
functional connectivity (Fig.~\ref{connillustr}b).
Synchrony is typically quantified by means of the synchronization index, also 
known as the phase locking value, see e.g., 
\cite{Tass-Rosenblum-Weule-Kurths-Pikovsky-Volkmann-Schnitzler-Freund-98,%
Rodriguez_et_al-99,Mormann-Lehnertz-David-Elger-00}. 
Since it is computed from the phases
of interacting units, see Eq.~(\ref{sind2}) below, it quantifies functional phase connectivity. 

Now we discuss the notion of \textit{effective phase connectivity}, crucial for
our approach.
For this purpose we recall, that if the coupling between the nodes is weak then
the dynamics 
of $N$ coupled systems can be reduced to that of $N$ phases. It means that
each oscillator, which can 
be a high-dimensional dynamical system, can be characterized by a single
variable, the phase $\vp_k$
\cite{Kuramoto-84,Pikovsky-Rosenblum-Kurths-01}. This variable is introduced in
such a way, that 
it grows uniformly in time, i.e.
\begin{equation}
\dot\vp_k=\w_k \;.
\label{autphase}
\end{equation}
The weakness of coupling means that the motion in the state space takes place on
 the 
smooth invariant $N$-dimensional torus. The equations of the phase dynamics are:
\begin{equation}
\dot\vp_k=\w_k +h_k(\vp_1,\vp_2,\ldots,\vp_N)\;,\quad k=1,\ldots,N\;,
\label{pheq}
\end{equation}
where the new phase coupling functions $h_k$ can be obtained from the state
space coupling functions
$\mathbf{H}_2(\mathbf{x}_1,\mathbf{x}_2,\ldots,\mathbf{x}_k)$ by means of a 
perturbative reduction  \cite{Kuramoto-84}
in the form of the power series in the coupling parameter $\e$: 
\begin{equation}
h_k(\vp_1,\vp_2,\ldots,\vp_N)=\e h_k^{(1)}(\vp_1,\vp_2,\ldots,\vp_N)+
\e^2h_k^{(2)}(\vp_1,\vp_2,\ldots,\vp_N)+\ldots 
\label{eq:power}
\end{equation}
If the coupling function $h_k$ depends on $\vp_l$, we say that there exist a
directed  phase connectivity link $l\to k$.

The crucial issue is that even if the structural connections are pairwise,
i.e. 
\begin{equation*}
\mathbf{H}_k=\sum_{j\ne k}\mathbf{H}_{kj}(\mathbf{x_k,x_j}) \;,
\end{equation*}
the phase coupling functions $h_k$ can have terms, \textit{depending on many
phases, not only on the phases of 
directly coupled nodes} \cite{Kralemann-Pikovsky-Rosenblum-11}. 
So, for the example illustrated in Fig.~\ref{connillustr}a, the equation of the 
phase dynamics for the third node has the form
\begin{equation*}
\dot\vp_3 = \w_3 +\e h_3^{(1)}(\vp_2,\vp_3) +\e^2 h_3^{(2)} (\vp_1,\vp_2,\vp_3) +\ldots\;.
\end{equation*}
Notice, that the indirect coupling  $1\to 2\to 3$ is reflected by the term $\sim\e^2$; 
hence, for a very weak coupling the indirect link $1\to 3$ is negligible and the effective phase
connectivity coincides with the structural one. 
However, for not very small $\e$, the indirect coupling  $1\to 3$ becomes not small either.
Notice, that the indirect coupling $3\to 2\to 1$ is not possible, since there is no information flow 
from  node $3$ to node $2$. 

Thus, generally, structural and effective phase connectivities of a network
differ (Fig.~\ref{connillustr}c), although for a small coupling they nearly coincide. 
Below we present a technique for reconstruction of effective phase
connectivity from multichannel data.

\section{Effective connectivity from phase dynamics}
Suppose the phase model Eq.~(\ref{pheq}) of the system of $N$ interacting
self-sustained oscillators is known.
It is convenient to represent the right hand side of Eq.~(\ref{pheq}) as a
Fourier series:
\begin{equation}
\hspace*{-7mm}\w_k +h_k(\vp_1,\vp_2,\ldots,\vp_N) = 
\sum_{l_1,\ldots,l_N}
\mathcal{F}^{(k)}_{l_1,\ldots,l_N}\exp{[\rmi(l_1\vp_1+l_2\vp_2+\ldots+l_N\vp_N)]
}\;.
\label{pheqrhs}
\end{equation}
The norm of the coupling function $h_k$ quantifies influence of the rest of
the 
network on the oscillator $k$.
In order to quantify the action of a particular oscillator $j$, we introduce the
\textit{partial norms} 
\begin{equation}
\mathcal{N}_{k\leftarrow j}^2= 
\sum_{l_k,l_j\ne 0} \left |
\mathcal{F}^{(k)}_{0,\ldots,l_k,0,\ldots,l_j,0,\ldots} \right |^2 \;,
\label{pnorm}
\end{equation}
i.e. from the Fourier series we pick up only the terms depending on the phases
$\vp_k$, $\vp_j$ and terms, depending on  $\vp_j$.
Notice that in a similar way, picking up  terms depending on three phases, 
one can quantify the joint action of two oscillators on the chosen one, 
$m,j\to k$, etc, see \cite{Kralemann-Pikovsky-Rosenblum-11}.

Thus, if the phase model (\ref{pheqrhs}) can 
be reconstructed from data, we can compute partial
norms 
$\mathcal{N}_{k\leftarrow j}$ for all combinations $k,j$ and in this way obtain
a description 
of network connectivity. This description is quantitative and directional,
since generally  
$\mathcal{N}_{k\leftarrow j}\ne \mathcal{N}_{j\leftarrow k}$.

\section{Phase dynamics from data}
\subsection{General approach}
The first step of the approach is to obtain phases of all oscillators.
We assume that we measure the outputs of all nodes of the network and that these
outputs are 
suitable for phase estimation. 
Hence, we start with $N$ oscillatory time series and 
then compute $N$ continuous-time protophases $\theta_k(t)$
\cite{Kralemann_et_al-07,Kralemann_et_al-08}, using e.g., the Hilbert Transform 
or any other embedding. (The choice of the optimal technique for protophase
estimation is 
beyond the framework of the current study.)
We recall, that the true phase of any autonomous self-sustained oscillator shall
satisfy Eq.~(\ref{autphase}). 
The angle variables (protophases) $\theta_k$ provided by the Hilbert transform
or other techniques 
generally do not have this property and therefore shall be transformed to true
phases 
$\theta_k(t)\to\vp_k(t)$, as discussed in the~\ref{app_proto}.

Next step is to compute numerically the derivatives of all phases (here it is
done simply 
by computing central final difference; for noisy data the application of the
Savitzky-Golay filter 
is recommended) and to fit the coefficients of the model (\ref{pheq}), as
discussed in 
\cite{Kralemann-Pikovsky-Rosenblum-11}. 

As discussed below, reconstruction is not possible if 
oscillators synchronize.
Practically, in order to exclude from the consideration the synchronous states,
we compute pairwise 
and triplet synchronization indices \cite{Kralemann-Pikovsky-Rosenblum-13} 
according to 
\begin{equation}
\gamma_{j,k}=\left | \langle \exp{[\rmi (n\vp_j-m\vp_k)]}\rangle \right | 
\label{sind2}
\end{equation}
and
\begin{equation}
\gamma_{j,k,l}=\left | \langle \exp{[\rmi (n\vp_j+m\vp_k+p\vp_l)]}\rangle \right
|\;. 
\label{sind3}
\end{equation}
Notice that in Eq.~(\ref{sind2}) $n,m$ are positive integers, while in
Eq.~(\ref{sind3}) $n,m,p$
can be both positive and negative. 

\subsection{Effect of network size and of synchrony level}

In the presented approach we face a numerical problem related to the system's dimension. 
Consider first $N=2$, then the coupling functions $h_{1,2}$ are functions of two 
variables $\vp_{1,2}$ and in order to recover them 
by means of fitting we need that the data points cover 
the square $0\leq\vp_12<\pi,\,0\leq \vp_2<2\pi$. 
If two oscillators synchronize (i.e. the index (\ref{sind2}) is close to one), 
then $\vp_1,\vp_2$ are functionally dependent, 
the points do not cover the square but fall on a line, 
and the function of two variables cannot be recovered. 
If the systems are close to synchrony, the distribution of the points in the square
is very inhomogeneous: the points mostly concentrate along a line but occasionally 
the trajectory makes an excursion, e.g., due to noise-induced phase slips.
In this case the reconstruction is possible if the observation time is sufficiently long.
A detailed analysis of a dependence of accuracy of reconstruction on the values of 
synchronization index (\ref{sind2}) and of interrelation between synchrony level and 
required data length will be presented elsewhere.
In the present contribution we concentrate on the principal problems of network reconstruction 
and avoid the case of strong synchrony  by excluding from the further analysis 
the states  with $\gamma_{j,k}>0.5$. 
The chosen threshold is rather arbitrary; numerical tests demonstrate that it ensures a robust 
model reconstruction.

Let now $N=3$ and suppose that oscillators are far from synchrony. 
(Practically we again set a threshold for the triplet synchronization 
index (\ref{sind3}) and analyze only the states with $\gamma_{j,k,l}<0.5$.)
For successful data reconstruction we need enough points to cover the cube
$0\leq\vp_j< 2\pi$. 
In dimension three this is still feasible, however, the amount of data required to fill the hyper-cube
in dimension $N$ grows rapidly with the dimension, making the reconstruction of 
high-dimensional coupling functions hardly possible already for $N>3$.
Thus, in dimensions larger than three only partial phase dynamics reconstruction
is possible.

\subsection{Partial phase dynamics}
The simplest and straighforward approach is to perform a pairwise
analysis of the network, reconstructing the phase dynamics for all pairs of nodes. 
In this way one does not reconstruct the full Eq.~(\ref{pheq}),
but fits the pairwise coupling functions $h_{kj}$ to satisfy the equation
\begin{equation}
\dot\vp_k=\w_k + \sum_{j\neq k} h_{kj}(\vp_j,\vp_k)\;,
\label{pheq2}
\end{equation}
where all time series except for $\vp_k(t),\vp_j(t)$ are ignored.
The norms $\mathcal{P}_{k\leftarrow j}=||h_{kj}||$ are then used
for quantification of network connectivity.
As we show in detail in Section~\ref{sec:pta} below, this estimation 
may yield spurious effective phase connections. Indeed, suppose one has a
network 
like in Fig.~\ref{connillustr}a and determines phase connectivity according
to~\eqref{pheq2}. 
Because $\vp_1$ is correlated with $\vp_2$, and $\vp_2$ drives $\vp_3$,
consideration of the
pair $\vp_1,\vp_3$ will give non-zero coupling from $\vp_1$ to $\vp_3$. Only
full phase dynamics
according to Eq.~(\ref{pheq}) would reveal absence of direct coupling between
$\vp_1$ and $\vp_3$ and presence of indirect coupling $1\to 2 \to 3$.

Making use of this observation, we suggest to perform \textit{partial triplet analysis}
for networks with more than three oscillators as well.
In doing this, we reconstruct the three-dimensional coupling functions for 
all possible triplet configurations. 
Namely, from the time series of three phases
$\vp_j,\vp_k,\vp_l$ (all indices $j,k,l$ are different) 
we first reconstruct the coupling terms $h_{jkl},h_{klj},h_{ljk}$, 
ignoring all other phases.  
From these functions we compute, according to Eq.~(\ref{pnorm}), 
partial norms
$\tilde\mathcal{T}_{j\leftarrow k}(l)$ 
for all binary connections within this triplet 
(all these norms for a triplet correspond to different 
permutations of symbols $j,k,l$). 
Next, we repeat this procedure for all triplets within the network. 
As a result, we obtain $N-2$ estimates of $\tilde\mathcal{T}_{j\leftarrow k}$ 
for the connection $j\leftarrow k$, since the oscillators $j$ and $k$ belong to $N-2$ 
different triplets. 
We suggest to take the minimal value of these estimates as the final
triplet-based measure of the binary effective phase connectivity:
\begin{equation} 
\mathcal{T}_{j\leftarrow k}=\min_{l}\tilde\mathcal{T}_{j\leftarrow k}(l)\;.
\label{trest}
\end{equation}
To support this approach, let us assume that the motif in Fig.~\ref{connillustr}a is a part
of a larger network. From the triplets including oscillators $1$, $3$, and some disconnected 
oscillator $n$ one would obtain a strong binary term $3\leftarrow 1$. However, in the triplet
$\{123\}$ this term will be small, as here the analysis recognizes that in fact oscillator $3$ is 
driven by oscillator $2$, and not by oscillator $1$. This triplet will correctly yield a 
small value for the link $3\leftarrow 1$.

\section{Pairwise analysis vs triplet analysis}
\label{sec:pta}
\subsection{Three oscillators}

To demonstrate the method, we start with the case of three coupled units, where the
triplet analysis gives a full 
description of the phase dynamics, and compare the results with the partial pairwise
analysis.
We introduce our approach using the example of three coupled van der Pol
oscillators:
\begin{equation}
\eqalign{
\ddot x_1 -\mu(1-x_1^2)\dot x_1 +\w_1^2 x_1 = 
        \e [\sigma_{12}(x_2+\dot x_2) + \sigma_{13}(x_3+\dot x_3) ]  \;,\cr
\ddot x_2 -\mu(1-x_2^2)\dot x_2 +\w_2^2 x_2 = 
       \e [\sigma_{21}(x_1+\dot x_1) + \sigma_{23}(x_3+\dot x_3) ]  \;,\cr
\ddot x_3 -\mu(1-x_3^2)\dot x_3 +\w_3^2 x_3 = 
       \e [\sigma_{31}(x_1+\dot x_1) + \sigma_{32}(x_2+\dot x_2) ]  \;.\cr
}\label{vdp3} 
\end{equation}
Here the coefficients $\sigma_{ij}$ are either zero or one; they determine the
structure of the coupling network
in the three-oscillator system, whereas the parameter $\e=0.2$ determines the
coupling strength.
The other parameters are: $\mu=0.5$, $\w_1=1$,  $\w_2=1.3247$,  $\w_3=1.75483$.
The choice of the frequencies is motivated by an attempt to avoid synchronous
states, 
where the reconstruction is not possible.
We generate three-channel data by solving Eq.~(\ref{vdp3}) with time 
step $0.05$ and use $10^5$ 
data points per channel for the phase dynamics reconstruction;
this number of points corresponds to $\approx 800$ periods of the slowest oscillator $1$
 and to $\approx 1400$ periods of the fastest oscillator $3$. 
 The effect of the data length is analyzed in detail in \ref{app:num}.
The protophases $\theta_{1,2,3}$ are computed via the Hilbert transform
of observables $x_k$ and then
transformed to 
genuine phases $\vp_{1,2,3}$ via the phase transformation, see
~\ref{app_proto}. 
Next, we estimate coupling functions, approximated by a Fourier series, 
cf.~Eq.~(\ref{pheqrhs}), and compute partial norms 
$\mathcal{N}_{k\leftarrow j}$ according to Eq.~(\ref{pnorm}). 
Practically,   we use the Fourier series of order $5$. (This number
appears sufficient for the van der Pol oscillators with parameters used, as adding
Fourier modes does not practically change the results; for more relaxation oscillations
one should include more modes in the analysis.)
For comparison, 
we also perform pairwise analysis, i.e. we compute the coupling function for the
pair $k,j$ 
neglecting the third oscillator to obtain
${\mathcal{P}}_{k\leftarrow j}$.

\begin{figure}[ht!]
\centerline{\includegraphics[width=0.5\textwidth]{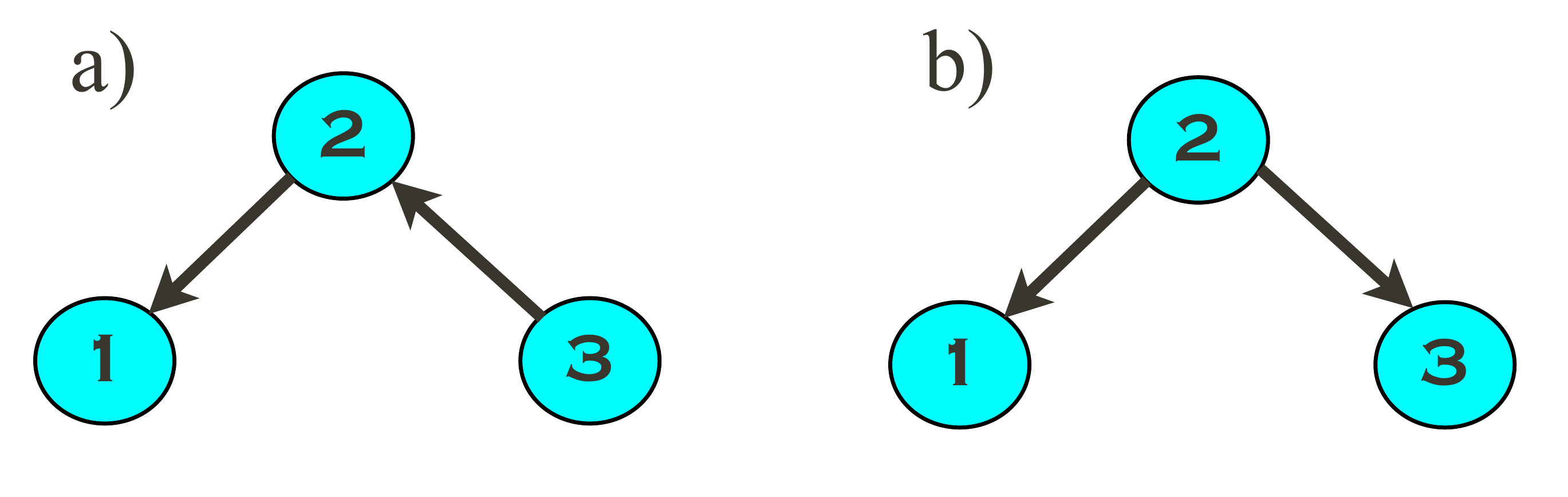}}
\caption{Two examples of three-oscillator networks. 
}
\label{connillustr2}
\end{figure}

We consider two examples of three-oscillator networks, illustrated in
Fig.~\ref{connillustr2}.
The corresponding results are given in Tables~\ref{tab1}(a,b), which
represent the 
reconstructed coupling matrices.

\begin{table}
\caption{\label{tab1}(a) Reconstruction of the coupling for the example in
Fig.~\ref{connillustr2}a.
An entry in the $k$-th row and $j$-th column represents
$\mathcal{N}_{k\leftarrow j}$,  
${\mathcal{P}}_{k\leftarrow j}$, which are measures of the coupling
strength.
With bold font we show existing structural connections. 
Hence, the corresponding entries should be much larger than the others; 
for the example under consideration this is obviously true. 
Notice that since the norms are by definition positive, we always obtain
overestimated values 
for non-existing connections. (b) Same as (a), but for the network
configuration shown 
in Fig.~\ref{connillustr2}b.\\[0.5ex]
}
\begin{indented}
\item[]\begin{tabular}{|c|c|c|c|}
\hline 
\rule[-1ex]{0pt}{2.5ex} \textbf{(a)} & $Osc_1$ & $Osc_2$ & $Osc_3$ \\ 
\hline 
\rule[-1ex]{0pt}{2.5ex} $Osc_1$ & • &  \textbf{0.103 , 0.104} & 0.018 , 0.024
\\ 
\hline 
\rule[-1ex]{0pt}{2.5ex} $Osc_2$ & 0.002 , 0.009 & • & \textbf{0.095 , 0.095} \\ 
\hline 
\rule[-1ex]{0pt}{2.5ex} $Osc_3$ & 0.001 , 0.001 & 0.001 , 0.001 & • \\ 
\hline 
\multicolumn{4}{c}{} \\
\hline 
\rule[-1ex]{0pt}{2.5ex} \textbf{(b)} & $Osc_1$ & $Osc_2$ & $Osc_3$ \\ 
\hline 
\rule[-1ex]{0pt}{2.5ex} $Osc_1$ & • & \textbf{0.113 , 0.113} & 0.003 , 0.016 \\ 
\hline 
\rule[-1ex]{0pt}{2.5ex} $Osc_2$ & 0.001 , 0.001 & • & 0.001 , 0.001 \\ 
\hline 
\rule[-1ex]{0pt}{2.5ex} $Osc_3$ & 0.005 ,  0.020 & \textbf{0.092 , 0.092} & •
\\ 
\hline 
\end{tabular} 
\end{indented}
\end{table}
 
First we discuss the results for the chain of oscillators
(Fig.~\ref{connillustr2}a), see Table~\ref{tab1}(a). 
We emphasize several important issues. (i) The reconstructed measures of the
coupling strength, i.e. 
the norms $\mathcal{N}_{k\leftarrow j}$,  
${\mathcal{P}}_{k\leftarrow j}$, corresponding to the existing
structural 
connections are almost two orders of magnitude larger than the norms for
unconnected oscillators, with an exception for the link 
$1\leftarrow 3$, discussed next.  
(ii) While oscillator $3$ is not connected 
to oscillator $1$ directly, there exist an indirect causal link mediated by
oscillator $2$.
Therefore the exceptionally large value $\mathcal{N}_{1\leftarrow 3}=0.018$
captures 
some real flow of information. Contrary, all values of the norms which
correspond neither 
to a direct nor to an indirect causal link, practically vanish. This fact
illustrates that 
although the effective phase connectivity generally differs from the structural
one, 
this difference is not due to artifacts, but reflects the properties of the
network.
(iii) As expected, the obtained norms are asymmetric, what shows that our
analysis captures 
essentially more than simple correlation.  
(iv)  While the results from the triplet and the pairwise analyses, i.e.
$\mathcal{N}_{k\leftarrow j}$ and  
${\mathcal{P}}_{k\leftarrow j}$, practically coincide for existing
structural links, 
they differ for some absent structural connections. 
Here, the values from the triplet analysis are notably smaller than those from
the pairwise one, 
cf. $\mathcal{N}_{2\leftarrow 1}=0.002$ and $\mathcal{P}_{2\leftarrow
1}=0.009$.
This happens because the oscillator 3 is not included into the pairwise model
and its 
effect on the correlation between oscillators 1 and 2 is erroneously assigned
to 
the link from 2 to 1. This essential difference between the pairwise and the triple
analysis will be illustrated in detail below.

In the second example (Fig.~\ref{connillustr2}b) two oscillators are driven by
the third one.
The results in Table~\ref{tab1}(b) corroborate those from the first example.
Indeed, the 
reconstructed norms for existing connections are again essentially larger than
those for the non-existing
connections. Moreover, here the analysis yields a perfect coincidence of the
effective phase  
connectivity and of the given structural one. Since in this 
configuration there are no indirect causal links, this fact substantiates the
conjecture 
that differences between effective phase and structural connectivities are due
to 
indirect coupling. 
The results from the triplet and the pairwise analysis are again identical in case of
existing structural links and
likewise differ when no direct structural coupling is present and the dynamics is
not autonomous, as can 
be seen for links $1\to 3$ and $3\leftarrow 1$. We see that here the pairwise
analysis again mistakenly 
takes correlations due to the common drive for causal interaction. Hence, for this
link the pairwise analysis 
fails, while the triplet analysis yields reasonable results.
 
\subsection{Four oscillators}
\begin{figure}[ht!]
\centerline{\includegraphics[width=0.35\textwidth]{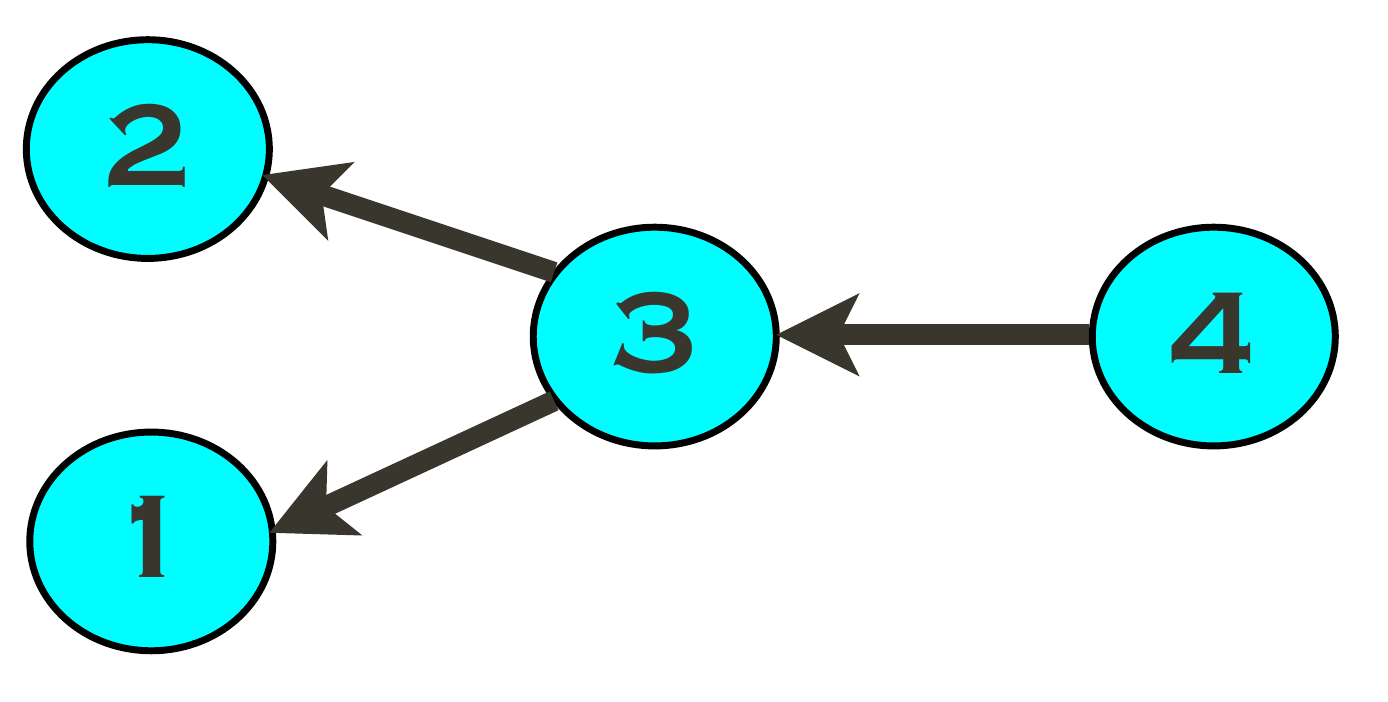}}
\caption{Four-oscillator network. 
}
\label{connillustr3}
\end{figure}

With the next example of a network (Fig.~\ref{connillustr3}) we want to address
the obvious problem:
for $N>3$ each oscillator enters more than one triplet configuration. 
So, e.g. the link  $1\leftarrow 2$ can be quantified by the analysis of the triplet 
configurations $\{123\}$ and $\{124\}$, and we have to choose between two
results.
To discuss this problem, we use the example of four coupled van der Pol
oscillators:
\begin{equation}
\eqalign{
\ddot x_1 -\mu(1-x_1^2)\dot x_1 +\w_1^2 x_1 = 
        \e [x_3+\dot x_3 ]  \;,\cr
\ddot x_2 -\mu(1-x_2^2)\dot x_2 +\w_2^2 x_2 = 
       \e [x_3+\dot x_3 ]   \;,\cr
\ddot x_3 -\mu(1-x_3^2)\dot x_3 +\w_3^2 x_3 = 
       \e [x_4+\dot x_4 ]  \;,\cr
\ddot x_4 -\mu(1-x_4^2)\dot x_4 +\w_4^2 x_4 = 
       0  \;,\cr
}\label{vdp4} 
\end{equation}
with  $\w_1=1$,  $\w_2=1.3247$,  $\w_3=1.75483$, 
and $\w_4=1.5333$.  All other parameters of the system, of the numerical
simulation, and of the data processing are
the same as in the previous examples. The results are given in Table~\ref{tab3}.
The values in the last row are due to numerical errors;  indeed, since
oscillator 4 is autonomous, 
these values should be zero. However, they are much less than the values
corresponding to structural connections.
The analysis of other entries supports our conjecture that the value of
$\mathcal{T}_{k\leftarrow j}$ is overestimated, 
if an oscillator which drives  the unit $k$ is excluded from the triplet. The
reason is that the correlations between the 
oscillators $k,j$ due to an external (with respect to the considered triplet)
drive are spuriously explained as a coupling 
between  $k,j$. Consider, e.g., the values of $\mathcal{T}_{1\leftarrow 4}$: in
the triplet configuration $\{134\}$ which 
includes both units driving the first oscillator, $\mathcal{T}_{1\leftarrow
4}\approx 0.016$, i.e. about two times smaller 
than in the configuration $\{124\}$.  Similar observation can be done for the
link $2\leftarrow 4$. For a comparison we also present the values obtained from the 
pairwise analysis; for the absent links they are overestimated as well.

\begin{table}
\caption{\label{tab3}
Reconstruction of the coupling for the example in
Fig.~\ref{connillustr3}.
An entry in the $k$-th row and $j$-th column represents two values for the
partial norm,  
$\tilde\mathcal{T}_{k\leftarrow j}$,  obtained from two different triplets which
include nodes $k,j$;
the corresponding triplets are denoted as $\{kjn\}$, the values obtained from the pairwise
analysis are denoted as $\{kj\}$. 
Existing structural connections are shown with bold font. The minimal coupling norms
$\mathcal{T}_{k\leftarrow j}$ are shown in boxes. 
Notice that the coupling norms for the truly 
existing connections are typically 80 to 100 times larger than those for the
truly non-existing connections.\\[0.5ex]}
\begin{indented}
\item[]\begin{tabular}{|c|c|c|c|c|}
\hline 
\rule[-1ex]{0pt}{2.5ex} • & $Osc_1$ & $Osc_2$ & $Osc_3$ & $Osc_4$ \\ 
\hline 
\rule[-1ex]{0pt}{2.5ex} $Osc_1$ & • & \fbox{0.002} \{123\} & \textbf{0.095} \{123\}  &
0.034 \{124\} \\ 
\rule[-1ex]{0pt}{2.5ex} • & • & 0.013 \{124\} & \fbox{\textbf{0.093}} \{134\} & \fbox{0.016}
\{134\} \\ 
\rule[-1ex]{0pt}{2.5ex} • & • & 0.016 \{12\} & \textbf{0.095} \{13\} & 0.033
\{14\} \\
\hline 
\rule[-1ex]{0pt}{2.5ex} $Osc_2$ & \fbox{0.003} \{123\} & • & \textbf{0.092}  \{123\} & 
0.046 \{124\}\\ 
\rule[-1ex]{0pt}{2.5ex} • & 0.007 \{124\} & • &  \fbox{\textbf{0.088}} \{234\} & \fbox{0.016}
\{234\} \\ 
\rule[-1ex]{0pt}{2.5ex} • & 0.009 \{12\} & • &  \textbf{0.092} \{23\} & 0.043
\{24\} \\
\hline 
\rule[-1ex]{0pt}{2.5ex} $Osc_3$ & \fbox{0.003} \{123\} & 0.010  \{123\} & • &
\fbox{\textbf{0.100}} \{134\} \\ 
\rule[-1ex]{0pt}{2.5ex} • & 0.009 \{134\} &  \fbox{0.009} \{234\} & • & \textbf{0.100}
\{234\} \\
\rule[-1ex]{0pt}{2.5ex} • & 0.003 \{13\} &  0.010 \{23\} & • & \textbf{0.100}
\{34\} \\  
\hline 
\rule[-1ex]{0pt}{2.5ex} $Osc_4$ & \fbox{0.001} \{124\} & \fbox{0.001} \{124\} & \fbox{0.001} \{134\}
& • \\ 
\rule[-1ex]{0pt}{2.5ex} • &  0.001 \{134\} & 0.001 \{234\} & 0.001 \{234\} & •
\\ 
\rule[-1ex]{0pt}{2.5ex} • &  0.001 \{14\} & 0.001 \{24\} & 0.001 \{34\} & •
\\
\hline 
\end{tabular} 
\end{indented}
\end{table}

The results support the suggested strategy to analyze networks of oscillators by
covering them with all possible triplets.
In a network of $N$ oscillators each partial norm $\mathcal{T}_{k\leftarrow j}$
is then computed from $N-2$ triplets,  and, 
hence, $N-2$ different estimates of its value are available. 
Since improper triplets overestimate $\mathcal{T}_{k\leftarrow j}$,  we choose
the minimal over all triplets
value $\mbox{min} \lbrace \mathcal{T}_{k\leftarrow j}\rbrace$ for the estimate
of the directional coupling 
between oscillators $k,j$.

\section{Random oscillator networks}
In this section we report on the statistical analysis of the quality of reconstruction
of effective phase connectivity in random networks of $N=5$ and $N=9$ oscillators, using the method
presented above. The equations of the model read
\begin{equation}
\ddot x_k -\mu(1-x_k^2)\dot x_k +\w_k^2 x_k =\e \sum_l \sigma_{kl} 
        (x_l\cos\Theta_{kl}+\dot x_l\sin\Theta_{kl}) \;.
\label{vdp5} 
\end{equation}
For each run, the random frequencies $\w_k$ are taken from the uniform distribution $0.5<\w<1.5$.
The random asymmetric connection matrix $\sigma_{kl}$ is composed of zeros and ones, with a fixed
number of two incoming connections for each oscillator for $N=5$, and with four incoming
connections for $N=9$. Parameter $\Theta$ governing effective phase shift
of the coupling is taken from a uniform distribution $0\leq \Theta<2\pi$. The protophase estimates
are obtained as $\theta_k=\arctan(\dot x_k/\omega_k x)$ and then
transformed to genuine phases. From the latter ones we calculated the maximum over 
binary and triplet synchronization indices according to Eqs.~(\ref{sind2},\ref{sind3}) and analyzed only cases
with the maximal index less than $0.5$; for computation of the 
indices we used $|n|,|m|,|p|\le 5$.

\begin{figure}[ht!]
\centerline{\includegraphics[width=\textwidth]{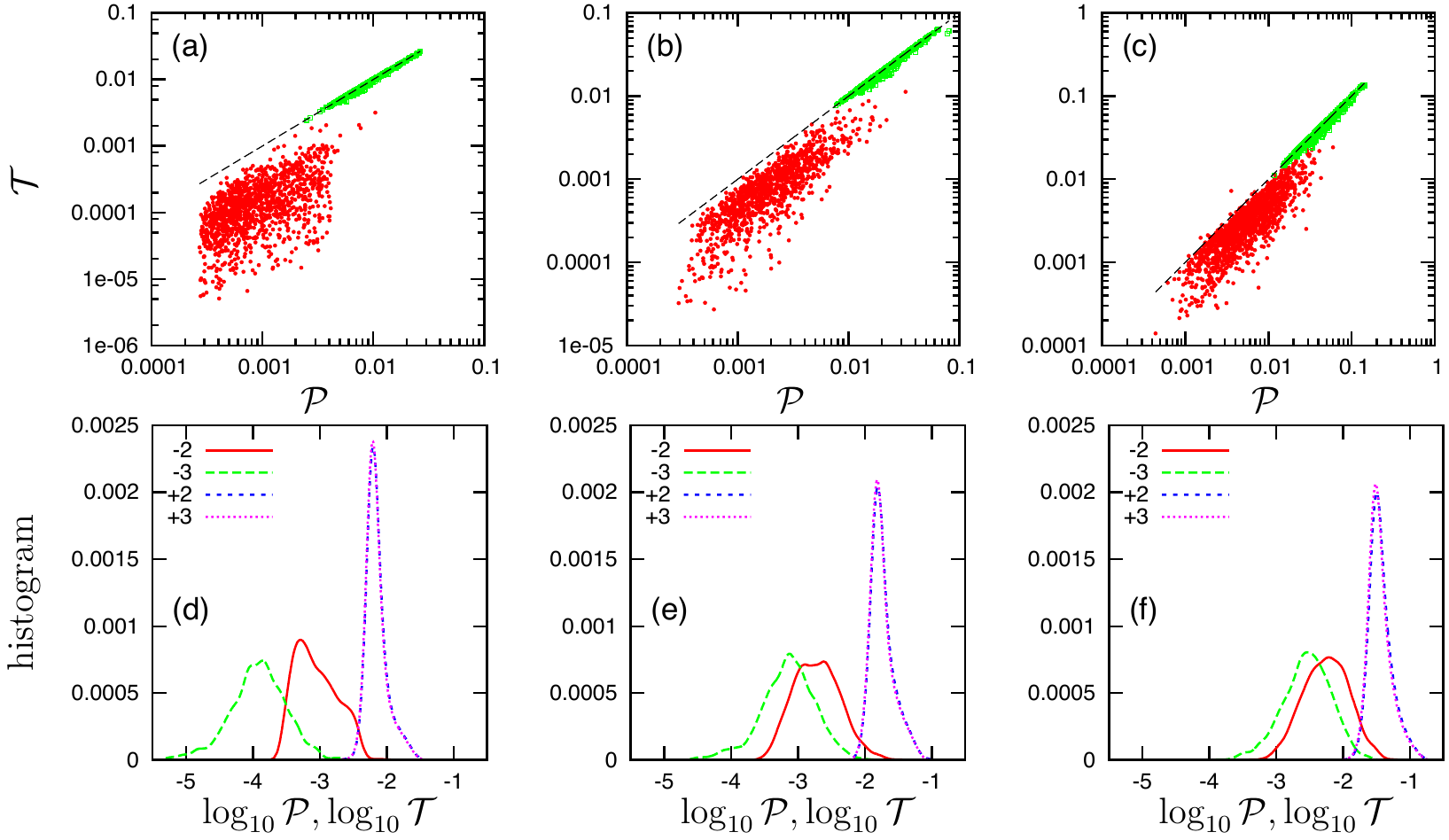}}
\caption{Results for random five-oscillator networks, for $\e=0.02$ (a,d), $\e=0.05$ (b,e), and 
$\e=0.1$ (c,f). In (a,b,c) the points, corresponding to existing and non-existing connections are 
shown with green and red, respectively; dashed line is the identity line.
Panels (d,e,f) show distributions for the norms ${\cal P}$ (red for non-existing and blue
for existing connections, also marked as $\mp 2$) and ${\cal T}$ 
(green for non-existing and magenta for existing connections, also marked as $\mp 3$).
For existing connections the distributions for ${\cal P},{\cal T}$ practically coincide,
while for non-existing connections they are essentially different: the distributions for 
the norms ${\cal T}$ obtained from the triplet analysis demonstrate very good separation between 
existing and non-existing connections. 
We quantify this separation by misclassification rates; the decision threshold is set to equalize 
errors of type I (non-existing connection is classified as existing) and of type II 
(existing connection is classified as non-existing). 
For cases (a,b,c), these rates  are respectively 
0.01, 0.0213, 0.0345 for the pairwise analysis and 0.0015, 0.0037, 0.0086 for the triplet analysis.
}
\label{fig:5rand}
\end{figure}

The results for $N=5$ and for three different values of $\e$ are presented in 
Fig.~\ref{fig:5rand}. Here we show the pairwise coupling norms $\mathcal{P}$ 
vs the coupling norms $\mathcal{T}$, based on the triplet analysis, for the same data. 
Norms corresponding to the existing couplings ($\sigma_{kl}=1$) 
are marked with green, while norms corresponding to non-existing connections ($\sigma_{kl}=0$) are shown
with red; $10^6$ points are used for the analysis. 
Summarizing these results, we conclude: (i) For the existing connections, the
triplet analysis practically does not change the value of the partial norm, obtained from the pairwise
analysis: all green points lie very close
to the diagonal (dashed line), so that $\mathcal{P}\approx \mathcal{T}$. (ii) For non-existing 
connections, the triplet analysis significantly reduces the values of the
coupling norms. For small coupling
strength $\e=0.02$ this reduction is rather strong, with typical factor about $0.3$. For larger 
coupling the  reduction is less pronounced. (ii) The values of the 
coupling from the pairwise analysis
do not allow to distinguish between existing and non-existing connections unambiguously, 
as the distributions 
of the obtained norms overlap (panels (d,e,f), 
cf. similar observation in~\cite{Kralemann-Pikovsky-Rosenblum-11}). 
However, for the norms obtained from the triplet analysis, the existing and non-existing 
couplings can 
be clearly separated, at least for the small coupling strength $\e=0.02$.
The separation remains non-perfect for larger coupling strengths, 
but nevertheless it is definitively better than in case of the pairwise analysis. 

\begin{figure}[ht!]
\centerline{\includegraphics[width=0.5\textwidth]{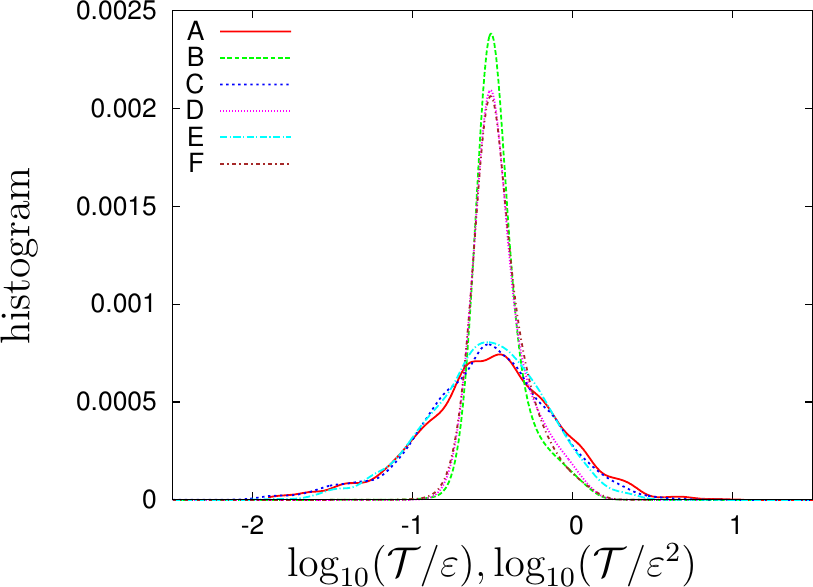}}
\caption{Scaling of coupling terms for random five-oscillator networks. 
Here the distributions of ${\cal T}$ from Fig.~\ref{fig:5rand}d,e,f are re-scaled: 
the norms of existing links are divided by $\e$, whereas the 
norms of non-existing connections are divided by $\e^2$. 
The result confirms our conjuncture that terms, describing indirect phase connectivity appear
in the second order of the phase approximation, i.e. are $\sim \e^2$. 
Moreover, it confirms that most of the non-existing connections revealed by the technique 
are not artifacts, but reflect causal information flow, mediated by the indirect driving.  
Curves A,B correspond to the rescaled
distributions marked as $\mp 3$ in Fig.~\ref{fig:5rand}d, while
C,D and E,F are the corresponding re-scaled distributions from 
Fig.~\ref{fig:5rand}e and Fig.~\ref{fig:5rand}f,
respectively. 
}
\label{fig:5rand_rescaled}
\end{figure}

In our approach we relied on Eq.~(\ref{eq:power}) and on the statement that the 
indirect connections are reflected by the Fourier terms of the coupling function,
proportional to $\e^2$, while the corresponding terms for the structurally existing 
links appear already in the first approximation in $\e$. 
This is confirmed by Fig.~\ref{fig:5rand_rescaled}, where we present the re-scaled 
distributions of norms  $\mathcal{T}$. The overlap of re-scaled distributions firmly supports
that the effective indirect phase connectivity is not the artifact of the method, but a
real coupling appearing in the second order in the coupling strength.

\begin{figure}[ht!]
\centerline{\includegraphics[width=0.6\textwidth]{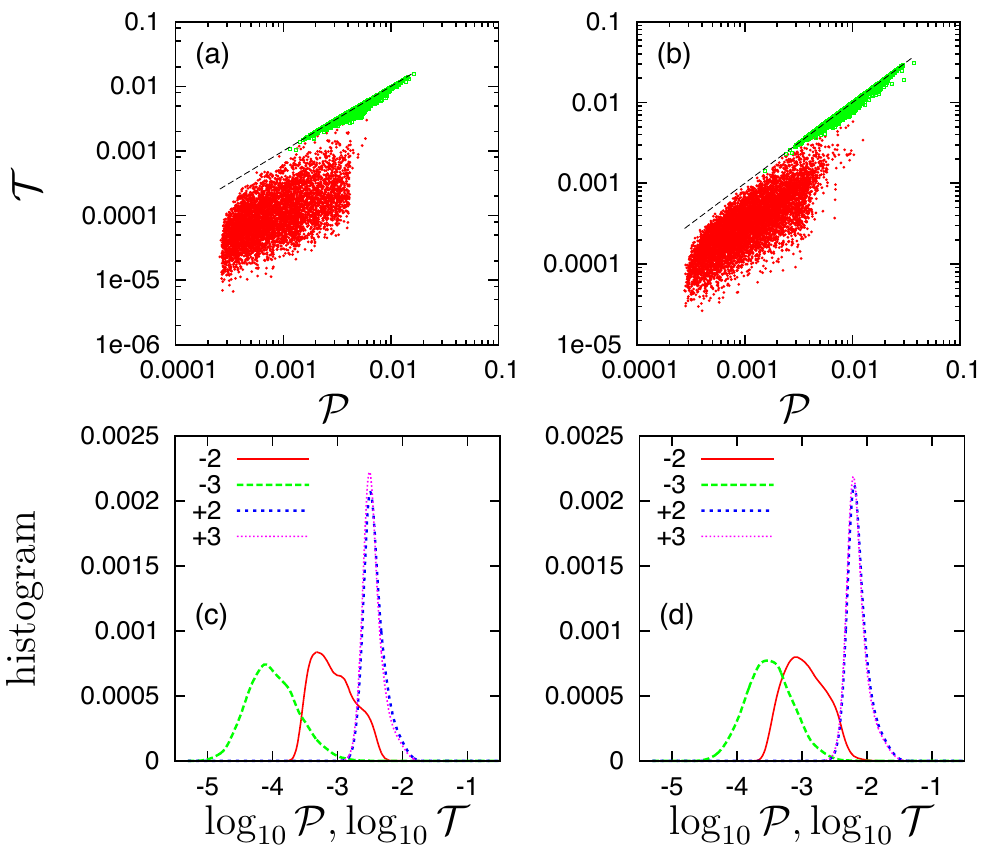}}
\caption{Same as Fig.~\ref{fig:5rand} but for nine-oscillator networks,  
for $\e=0.01$ (a,c) and $\e=0.02$ (b,d). 
}
\label{fig:9rand}
\end{figure}

\begin{figure}[ht!]
\centerline{\includegraphics[width=0.5\textwidth]{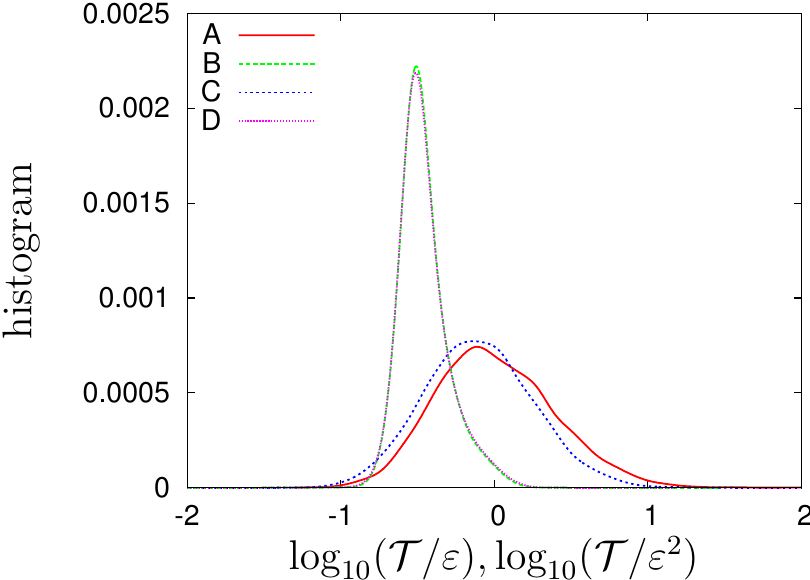}}
\caption{Same as Fig.~\ref{fig:5rand_rescaled} but for random 9-oscillator networks. 
Curves A,B and C,D correspond to distributions marked as $\mp 3$ in Fig.~\ref{fig:9rand}c
and Fig.~\ref{fig:9rand}d, respectively.
}
\label{fig:9rand_rescaled}
\end{figure}

The results for nine-oscillator networks, are shown in 
Figs.~\ref{fig:9rand},\ref{fig:9rand_rescaled}, 
for two values of the coupling strength; here $10^5$ points are used for the analysis.
The results are generally similar to those for five coupled oscillators.
Notice the shift of maxima of curves A,C in Fig.~\ref{fig:9rand_rescaled} with respect to
those of curves B,D; this reflects the fact that our approach always 
overestimates the norms. Since the coupling in this example is rather weak (because of a larger
number of incoming links compared to five-oscillator networks), the norms 
proportional to $\sim \e^2$ are of the order of numerical precision. However,
as follows from Figs.~\ref{fig:9rand}c,d, the separation between existing and non-existing 
links is still very good.

\section{Discussion}

\subsection{Relation to other techniques}
A large variety of different techniques, aimed at the quantification of 
interdependencies between time series, have been presented in the literature.
A detailed comparison of the performance of these methods is a 
serious computational and programming 
task, and requires a separate study; see, e.g., 
\cite{PhysRevE.71.036207,Kreuz200729,Osterhage_et_al-08,%
Silfverhuth_et_al-12,Greenblatt-Pflieger-Ossadtchi-12}. 
Several algorithms are implemented 
in software packages \cite{Trentool1,Trentool2,Seth-10}, see also {\tt http://hermes.ctb.upm.es}.

Here we discuss briefly two other methods. The first one is based
on Granger causality and is implemented as a user-friendly 
toolbox tailored to evaluate the causal connectivity underlying neural 
data~\cite{Seth-10,Barnett-Seth-14}.
With this toolbox we estimated the coupling structure for the four coupled van der Pol oscillators, 
Eqs.~(\ref{vdp4}). Comparison with our method, presented in details in \ref{app:cgc},
shows that both methods basically correctly estimate structural links.
On the one hand, our method provided more homogeneous results (all existing links have approximately
the same level of coupling, while the Granger causality approach gives couplings differing up
to $75 \%$). On the other hand, the phase model based results indicate weak indirect coupling for 
not directly linked oscillators, while the values of the Granger connectivity in this cases 
correspond exactly to the structural coupling. This comparison shows that quality of
our approach 
of revealing the connectivity of oscillatory networks is
comparable to the Granger causality approach, 
with an advantage that its results are widely insensitive to the choice of parameters.

Next, we discuss a relation to the technique \cite{Shelter_et_al-06}, based on computation of 
synchronization indices, cf. \cite{Smirnov-Schelter-Winterhalder-Timmer-07}. Since the indices are
a symmetric measure, this technique yields a non-directed measure of phase interdependence, 
i.e. it quantifies functional phase connectivity. The method also differentiates between existing 
and non-existing links. Reliable performance requires that synchronization indices are rather 
high, so that the system should be close to synchrony, and is based on the statistics
of the deviations from this synchrony (ideal synchrony yields identity matrix 
which contains no information). 
On the contrary, performance of our technique  is better
if the network is far from synchrony; thus, in this respect the methods are complimentary.
However, the method \cite{Shelter_et_al-06} requires the $1:1$ synchrony, i.e. closeness
of the frequencies of all nodes; our technique does not have this limitation.

As already discussed in Section~\ref{sec2} and illustrated in Fig.~\ref{connillustr}, 
the main aim of our approach
is to provide a measure for \textit{directional} effective phase connectivity. 
In this paper we present a further improvement of the previously reported 
studies~\cite{Rosenblum-Pikovsky-01,Kralemann_et_al-07,Kralemann_et_al-08,%
Kralemann-Pikovsky-Rosenblum-11} by including triplet interactions. 
This technique is complementary to other approaches where one characterizes 
\textit{non-directional} correlations, either
directly for processes or via synchronization indices of phases 
(see e.g.~\cite{Skudlarski-08,Bartsch26062012,%
Tass_et_al-98,Rodriguez_et_al-99,Axmacher2008,PhysRevE.73.031915,Kralemann-Pikovsky-Rosenblum-13}). 
In general, the phase dynamics reconstruction and correlation analysis can be applied to the same data sets, 
but they answer different questions, and therefore can be hardly compared directly. 
However, recently  correlation methods have been also adopted for 
reconstruction of directed networks~\cite{Levnajic-13,Levnajic-Pikovsky-14}. 
We believe that a synergetic application of different methods characterizing 
dynamical processes of network interactions among organ systems 
(see~\cite{Bashan_et_al-12,Ivanov-Bartsch-14} for recent advances)
would significantly contribute to progress in this field.

\subsection{Conclusions}

Discussing the pros and cons of our approach we emphasize, that it explicitly exploits the assumption that 
all nodes of the network are active, self-sustained oscillators, so that the
phase dynamics description 
is meaningful. This is a disadvantage of the approach, compared to 
the information theory based methods which are 
free of any assumptions regarding the type of the dynamics. 
On the other hand, for this particular case our 
technique yields an appropriate description which admits a clear physical interpretation.  
We clearly interpret the output of our algorithms and the recovered network structure, 
by theoretical argumentation and numerical demonstration,
as revealing the effective phase connectivity 
which coincides with the structural one in case of very weak coupling and shows additional links when 
the coupling is not too weak. Noteworthy, 
the additional links are not artifacts: most of them quantify the indirect 
influence between nodes appearing in the higher orders of phase reduction,
and correspond to a causal information flow. Although we cannot unambiguously 
separate the links which are physically meaningful from those which are due to computational errors, 
we suggest as a rule of thumb a practical approach based on the triplet analysis. 
Indeed, the computed norms are always positively biased and can never 
be zero. However, as shown in the examples of Fig.~\ref{connillustr2}a and Fig.~\ref{connillustr3}, see 
also \cite{Kralemann-Pikovsky-Rosenblum-11}, the norms corresponding to the real information flow
are always at least several times larger than the noise level. 
Figure~\ref{fig:5rand_rescaled} demonstrates,
that the width of the distribution for the norms of non-existing structural connections 
is about two orders
of magnitude. Hence, we can reasonably assume that norms that are, say, five times larger than 
the minimal
one, correspond to indirect connections. The discrimination can be improved using the information on 
the direct connections (which can be recovered reliably, as discussed above): if there is a path 
between nodes $j,k$ via node $m$, then the indirect connection $j\to k$ is very likely to exist.  
 
We stress that although the method is based on the phase description Eq.~(\ref{pheq}), it is not restricted 
to the case of very weak coupling. Indeed, \textit{analytical} derivation of Eq.~(\ref{pheq}) requires 
weakness of coupling, but numerical reconstruction is possible as long as the signals 
can be considered 
as nearly periodic or weakly chaotic. Certainly, application of the method implies 
that the outputs of all 
nodes, suitable for phase estimation, are available.

An important issue is that Eq.~(\ref{pheq}) is valid also for transient processes. 
Hence, the techniques does not imply stationarity of the data. So, e.g., if the network is repeatedly 
stimulated (cf.~\cite{Wagner-Fell-Lehnertz-10,Levnajic-Pikovsky-11}), 
the pieces of data between 
the stimuli can be used for the phase dynamics reconstruction and,
therefore, characterization of the network connectivity. Moreover, repeated 
perturbation can be a useful tool 
in case when the network (or some nodes) are close to synchrony, so that the reconstruction 
without perturbation fails. 
Perturbing the system and observing its relaxation to 
the synchronous state one can obtain enough data
for successful application of the technique.  This feature makes the technique
suitable for event-related analysis. Indeed, if a single evoked response is too short, 
one can use for averaging data obtained in several trials.

Although the proposed technique yields a good differentiation between 
existing and non-existing links, the determination of statistical significance
of small values of the connectivity measure requires an additional analysis.
(We remind that these values are always positive, thus one needs a threshold 
for discriminating artifacts.)  
A natural approach is to use
a surrogate data test. An example of such a test for the case of two
physiological oscillators is given in \cite{Kralemann_et_al-13}. This 
test can be potentially extended for triplets and small networks, although
a problem of constructing surrogates with a prescribed structure of cross-correlations
has to be solved.
Since our triplet-based method performs better than the pairwise technique
for the same amount of data,
we expect that this better performance will manifest itself in the surrogate 
test analysis, to be reported elsewhere. 
Besides, since we achieve smaller values for the truly absent connections,
simple ad hoc rules like neglecting all
connections which are weaker than a certain percent of the strongest one
become more reliable.

As an issue for further extension of the technique we mention accounting of terms, 
depending on the phases of three nodes. 
These terms appear in the second approximation in $\e$ even for a purely pairwise
coupling and can be of the order of $\e$, if the coupling 
is nonlinear \cite{Kralemann-Pikovsky-Rosenblum-11}. 
These terms describe the joint action of  two oscillators on the third one. 
Quantification of this action is straightforward,  but representation of results
for a network of more than three oscillators is not trivial.

In summary, we have shown that the triplet analysis of an oscillator network 
yields an essential improvement in the quantification of the connectivity compared to 
the conventional pairwise analysis. 
The described technique provides effective phase connectivity and allows one a reliable 
differentiation between structurally existing and  non-existing links. 

\appendix
\section{Protophase to phase transformation}
\label{app_proto}
The true phase $\vp$ of an autonomous oscillator is introduced according to
Eq.~(\ref{autphase}), i.e. it grows uniformly with time.
However, an angle variable, or a protophase $\theta$, obtained from a scalar time
series by means of a two-dimensional 
embedding, e.g. via the Hilbert transform, generally does not have this property
and obeys
\[
\dot\theta = \w+g(\theta)\;.
\]
If the oscillator is coupled, e.g. to another one, then its phase dynamics is
described by
\[
\dot\theta_1 = \w_1+g(\theta_1)+h(\theta_1,\theta_2)\;,
\]
where the coupling function $h$ -- the goal of our analysis -- is small while
the function $g$ is generally not,  cf. Eq.~(\ref{pheq}).
Hence, the reconstruction of  the function $h$ is hampered by the function $g$
and the latter shall be eliminated by means of the
transformation $\theta\to\vp(\theta)$. 
For a noise-free system the transformation is easily obtained from
Eq.~(\ref{autphase}) using the chain rule: 
$\frac{\rmd \vp}{\rmd \theta}\frac{\rmd \theta}{\rmd t}=\w$ yields 
\[
\frac{\rmd \vp}{\rmd \theta}=\frac{\w}{\dot \theta} \;, \qquad\mbox{or}\qquad
\vp=\w\int_0^\theta\frac{\rmd \theta}{\dot\theta}\;.
\]
For noisy systems we have to average $\frac{\rmd t}{\rmd\theta}(t)$ over the
trajectory to obtain the transformation function
$\sigma(\theta)$:
\[
\frac{\rmd \vp}{\rmd \theta} = \sigma(\theta) = \w \left .\left\langle
\frac{\rmd t}{\rmd\theta}(t)  \right\rangle \right |_\theta \;,
\]
Up to a factor $2\pi$, this function is the probability density of $\theta$ and,
hence, can be easily obtained 
from the time series $\Theta(t_k)$, $k=1,\ldots,M$, see
\cite{Kralemann_et_al-07,Kralemann_et_al-08}. 
The final transformation reads:
\begin{equation}
\vp=\int_0^\theta\sigma(\theta')\rmd\theta'=\theta+2\sum_{n=1}^\infty\mbox{Im}
\left [ \frac{S_n}{n}(\exp(in\theta)-1)\right ]\;,
\label{transform}
\end{equation}
where $S_n=M^{-1}\sum_{k=1}^M\exp(-in\Theta_k)$ are coefficients of the Fourier
expansion of $\sigma(\theta)$.
Practically, one has to truncate the Fourier series and compute some finite
number $K$ of terms in Eq.~(\ref{transform}).
An efficient algorithm for determination of the optimal $K$ was suggested by
C.~Tenreiro ~\cite{Tenreiro-11}. 
A Matlab code for the
protophase-to-phase transformation with implementation of the
Tenreiro optimization can be downloaded 
from {\tt www.stat.physik.uni-potsdam.de/$\sim$mros/damoco2}. 

We illustrate the importance of the transformation by the following example. We
consider two \textit{uncoupled}  Hindmarsh-Rose
neuronal oscillators:
\begin{eqnarray*}
\dot x_{1,2} = y_{1,2}-x_{1,2}^3+3x_{1,2}^2-z_{1,2}+I_{1,2}\;,\\
\dot y_{1,2} = 1-5x_{1,2}^2-y_{1,2} \;,\\
\dot z_{1,2} = 0.006[4(x_{1,2}+1.56)-z_{1,2}]\;,
\end{eqnarray*}
where $I_1=5$, $I_2=5.1$. For the chosen parameter values both systems exhibit
periodic spiking. 
Next, we compute the protophases via Hilbert transform of $x_{1,2}$. Finally,
using Eq.~(\ref{sind2}) we compute 
the synchronization index twice, from the
protophases and from the true phases.
Since the systems are uncoupled, 
the true value is zero. The index, obtained from the protophases yields an obviously spurious
value $\rho\approx 0.13$, while the 
estimation using the true phases after the transformation above 
yields a reasonable result, $\rho\approx0.02$. 

\section{Effect of the length of the time series on model reconstruction}
\label{app:num}
\begin{figure}[ht!]
\centerline{\includegraphics[width=0.9\textwidth]{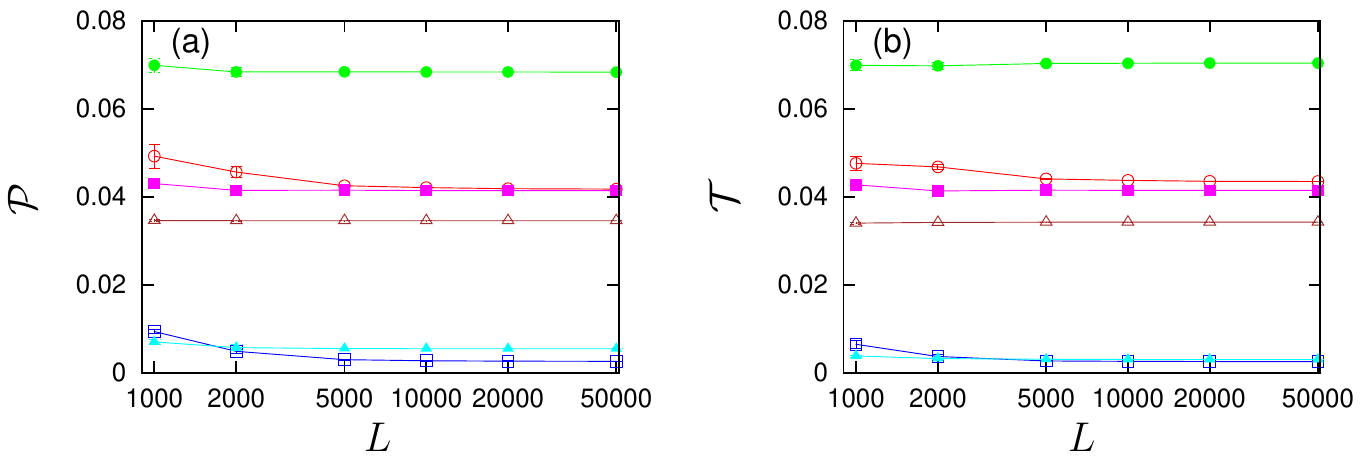}}
\caption{Dependence of the norms of coupling functions on the number $L$ of data 
points used for model reconstruction; the results of the pairwise and triplet analyses
are shown in (a) and (b), respectively. The symbols show the average 
values of ${\cal P}$ and ${\cal T}$ for all six structural  connections 
(four existing and two non-existing); the averaging is performed over 
100 runs with different initial conditions. 
Notice that the statistical variation is very small so that 
the error bars become invisible for large $L$.
}
\label{fig:npt}
\end{figure}

Here we discuss the accuracy of the model reconstruction on the 
number of data points and on the order of the Fourier series. 
For the test system we take the model (\ref{vdp3}) with  
$\sigma_{12} =\sigma_{13}=\sigma_{23}=\sigma_{31}=1$, 
$\sigma_{21}=\sigma_{32}=0$, i.e. there exist one bidirectional 
and two unidirectional connections. 
Other parameters are:  $\e=0.1$, $\mu=0.5$,  
$\w_1=1$,  $\w_2=1.3247$,  $\w_3=1.75483$.
Both pairwise and triplet analyses were performed for 
different number of points $L$.
The results illustrated in Fig.~\ref{fig:npt} clearly demonstrate 
that the network reconstruction is stable for $L\gtrsim 5000$.
Notice that this minimal data length corresponds to $\approx 40$
periods of the slowest oscillator.

\section{Comparison with Granger causality method}
\label{app:cgc}

Here we describe details of causal connectivity evaluation, based on Granger causality concept, 
with the help of the toolbox \cite{Seth-10}.
Namely, we reconstructed  the coupling pattern of four coupled van der Pol oscillators, see
Eqs.~(\ref{vdp4}), using the numerically obtained solutions $x_k(t)$ as an input for the toolbox. 
We followed the steps described in the manual, i.e. the signals have been detrended,  demeaned,
and standardized. Next, we have checked the statistical properties of the time 
series, namely whether the data are covariance stationary. 
We found, that neither the data nor their derivatives meet this requirement due to pronounced 
periodicity, which means that the Granger causality approach is
possibly not well-suited for our problem.
Thus, the results should be interpreted carefully. Evaluation of  the Granger connectivity from data
requires selection of a time lag parameter, $L$, which determines the order of an 
autoregressive model.
The algorithm for optimization of $L$ 
did not converge for our data. Therefore, we tried different parameter
values and found reasonable connectivity patterns for time lags $L=2$ and $L=3$. 
In both cases all values indicating causal influence were judged as 
significant results, see Table~\ref{tab:app:1}.   

\begin{table}[!hbt]
\centering
\caption{Results of application of the connectivity estimation based on Granger causality, for
two time lags $L=2$ and $L=3$ (shown in $\{\}$).}
\begin{tabular}{|c|c|c|c|c|}
\hline 
\rule[-1ex]{0pt}{2.5ex} • & $Osc_1$ & $Osc_2$ & $Osc_3$ & $Osc_4$ \\ 
\hline 
\rule[-1ex]{0pt}{2.5ex} $Osc_1$  & • & 0.000 \{2\} & \textbf{0.263} \{2\} & 0.000
\{2\} \\
\rule[-1ex]{0pt}{2.5ex} • & • & 0.000 \{3\} & \textbf{0.061} \{3\} & 0.006
\{3\} \\

\hline 
\rule[-1ex]{0pt}{2.5ex} $Osc_2$  & 0.000 \{2\} & • &  \textbf{0.113} \{2\} & 0.001
\{2\} \\
\rule[-1ex]{0pt}{2.5ex} • & 0.001 \{3\} & • &  \textbf{0.016} \{3\} & 0.010
\{3\} \\

\hline 
\rule[-1ex]{0pt}{2.5ex} $Osc_3$  & 0.000 \{2\} &  0.000 \{2\} & • & \textbf{0.065}
\{2\} \\
\rule[-1ex]{0pt}{2.5ex} • & 0.006 \{3\} &  0.001 \{3\} & • & \textbf{0.013}
\{3\} \\

\hline 
\rule[-1ex]{0pt}{2.5ex} $Osc_4$  &  0.000 \{2\} & 0.000 \{2\} & 0.000 \{2\} & •
\\
\rule[-1ex]{0pt}{2.5ex} • &  0.000 \{3\} & 0.000 \{3\} & 0.002 \{3\} & •
\\

\hline 
\end{tabular} 
\label{tab:app:1}
\end{table}

Although both results indicated a similar pattern of connectivity, the absolute values strongly 
varied with the time lag. Since the results for $L=2$ are closer to the true structural connectivity 
of the system, we compared them to our results. To simplify the comparison of different magnitudes,
we normalized the maximal values of the respective connectivity matrices to $1$. 
The values corresponding to structural coupling of the oscillators are shown with 
 bold font in Table~\ref{tab:app:2}. Ideally, 
this bold values should all be $1$ while all others should  be $0$. 
The values corresponding to structural coupling obtained from the phase model 
(the entries marked by PM) have less than $\sim 10\%$ deviance, while the results based on 
Granger causality (GC) exhibit an essential variation
(the difference between the maximal and minimal values is up to $75 \%$ of the maximal value);
recall that all structural connections in our simulation have same strength. 
On the other hand, the phase model based results indicate weak indirect coupling for 
not directly linked oscillators, while the values of the Granger connectivity in this cases 
correspond exactly to the structural coupling. 
This comparison shows that our approach is better suited for reconstructing the 
connectivity of oscillatory networks, because its results are widely insensitive to the 
choice of parameters.           
\begin{table}[!hbt]
\centering
\caption{Comparison of Granger connectivity method (data marked GC) with our phase dynamics 
method (data marked PM).}
\begin{tabular}{|c|c|c|c|c|}
\hline 
\rule[-1ex]{0pt}{2.5ex} • & $Osc_1$ & $Osc_2$ & $Osc_3$ & $Osc_4$ \\ 
\hline 
\rule[-1ex]{0pt}{2.5ex} $Osc_1$ & • & 0.02 \{PM\} & \textbf{0.93} \{PM\}  &
0.16 \{PM\} \\ 
\rule[-1ex]{0pt}{2.5ex} • & • & 0.00 \{GC\} & \textbf{1.00} \{GC\} & 0.00
\{GC\} \\ 

\hline 
\rule[-1ex]{0pt}{2.5ex} $Osc_2$ & 0.02 \{PM\} & • & \textbf{0.89}  \{PM\} & 
0.16 \{PM\}\\ 
\rule[-1ex]{0pt}{2.5ex} • & 0.00 \{GC\} & • &  \textbf{0.43} \{GC\} & 0.00
\{GC\} \\ 

\hline 
\rule[-1ex]{0pt}{2.5ex} $Osc_3$ & 0.03 \{PM\} & 0.09  \{PM\} & • &
\textbf{1.00} \{PM\} \\ 
\rule[-1ex]{0pt}{2.5ex} • & 0.00 \{GC\} &  0.00 \{GC\} & • & \textbf{0.25}
\{GC\} \\
  
\hline 
\rule[-1ex]{0pt}{2.5ex} $Osc_4$ & 0.01 \{PM\} & 0.01 \{PM\} & 0.01 \{PM\}
& • \\ 
\rule[-1ex]{0pt}{2.5ex} • &  0.00 \{GC\} & 0.00 \{GC\} & 0.00 \{GC\} & •
\\ 

\hline 
\end{tabular} 
\label{tab:app:2}
\end{table}

\section*{References}

\begin{thebibliography}{10}

\bibitem{10.1371/journal.pcbi.1000334}
J.W. Bohland, C.~Wu, H.~Barbas, H.~Bokil, M.~Bota, H.C. Breiter, H.T. Cline,
  J.C. Doyle, P.J. Freed, R.J. Greenspan, S.N. Haber, M.~Hawrylycz, D.G.
  Herrera, C.C. Hilgetag, Z.J. Huang, A.~Jones, E.G. Jones, H.J. Karten,
  D.~Kleinfeld, R.~K\"otter, H.A. Lester, J.M. Lin, B.D. Mensh, S.~Mikula,
  J.~Panksepp, J.L. Price, J.~Safdieh, C.B. Saper, N.D. Schiff, J.D.
  Schmahmann, B.W. Stillman, K.~Svoboda, L.W. Swanson, A.W. Toga, D.C.
  Van~Essen, J.D. Watson, and P.P. Mitra.
\newblock A proposal for a coordinated effort for the determination of
  brainwide neuroanatomical connectivity in model organisms at a mesoscopic
  scale.
\newblock {\em PLoS Comput Biol}, 5(3):e1000334, 03 2009.

\bibitem{Bullmore-Sporns-09}
E.~Bullmore and O.~Sporns.
\newblock Complex brain networks: graph theoretical analysis of structural and
  functional systems.
\newblock {\em Nature Reviews Neuroscience}, 10:187--198, 2009.

\bibitem{10.3389/fninf.2012.00014}
T.B. Leergaard, C.C. Hilgetag, and O.~Sporns.
\newblock Mapping the connectome: Multi-level analysis of brain connectivity.
\newblock {\em Frontiers in Neuroinformatics}, 6(14), 2012.

\bibitem{Boly_et_al-12}
M.~Boly, M.~Massimini, M.I. Garrido, O.~Gosseries, Q.~Noirhomme, S.~Laureys,
  and A.~Soddu.
\newblock Brain connectivity in disorders of consciousness.
\newblock {\em Brain Connectivity}, 2(1):1--10, 2012.

\bibitem{Pastrana-13}
E.~Pastrana.
\newblock Focus on mapping the brain.
\newblock {\em Nature Methods}, 10:481, 2013.

\bibitem{Sporns-13}
O.~Sporns.
\newblock Making sense of brain network data.
\newblock {\em Nature Methods}, 10(6):491--493, 2013.

\bibitem{10.1371/journal.pcbi.0010042}
O.~Sporns, G.~Tononi, and R.~K\"otter.
\newblock The human connectome: A structural description of the human brain.
\newblock {\em PLoS Comput Biol}, 1(4):e42, 09 2005.

\bibitem{Lehnertz-11}
K.~Lehnertz.
\newblock Assessing directed interactions from neurophysiological signals ---
  an overview.
\newblock {\em Physiological Measurement}, 32:1715--1724, 2011.

\bibitem{Mrowka_et_al-03}
R.~Mrowka, L.~Cimponeriu, A.~Patzak, and M.G. Rosenblum.
\newblock Directionality of coupling of physiological subsystems - age related
  changes of cardiorespiratory interaction during different sleep stages in
  babies.
\newblock {\em American J. of Physiology Regul. Comp. Integr. Physiol.},
  145:R1395--R1401, 2003.

\bibitem{Musizza_et_al-07}
B.~Musizza, A.~Stefanovska, P.~V.~E. McClintock, M.~Palu\v{s},
  J.~Petrov\v{c}i\v{c}, S.~Ribari\v{c}, and F.~Bajrovi\v{c}.
\newblock Interactions between cardiac, respiratory and {EEG}-delta
  oscillations in rats during an{\ae}sthesia.
\newblock {\em J Physiol}, 580(1):315--326, 2007.

\bibitem{Kralemann_et_al-13}
B.~Kralemann, M.~Fr\"uhwirth, A.~Pikovsky, M.~Rosenblum, T.~Kenner,
  J.~Schaefer, and M.~Moser.
\newblock In vivo cardiac phase response curve elucidates human respiratory
  heart rate variability.
\newblock {\em Nature Communications}, 4:2418, 2013.

\bibitem{Sharma_et_al-12}
S.~Das~Sharma, D.~S. Ramesh, C.~Bapanayya, and P.~A. Raju.
\newblock Sea surface temperatures in cooler climate stages bear more
  similarity with atmospheric {CO2} forcing.
\newblock {\em J. Geophys. Res.}, 117:D13110, 2012.

\bibitem{Wang_et_al-12}
G.~Wang, P.~Yang, X.~Zhou, K.~L. Swanson, and A.~A. Tsonis.
\newblock Directional influences on global temperature prediction.
\newblock {\em Geophys. Res. Lett.}, 39:L13704, 2012.

\bibitem{PhysRevLett.108.258701}
J.~Runge, J.~Heitzig, V.~Petoukhov, and J.~Kurths.
\newblock Escaping the curse of dimensionality in estimating multivariate
  transfer entropy.
\newblock {\em Phys. Rev. Lett.}, 108:258701, Jun 2012.

\bibitem{Sugihara26102012}
G.~Sugihara, R.~May, H.~Ye, Ch.-h. Hsieh, E.~Deyle, M.~Fogarty, and S.~Munch.
\newblock Detecting causality in complex ecosystems.
\newblock {\em Science}, 338(6106):496--500, 2012.

\bibitem{Skudlarski-08}
P.~Skudlarski, K.~Jagannathan, V.~D. Calhoun, M.~Hampson, B.~A. Skudlarska, and
  G.~Pearlson.
\newblock Measuring brain connectivity: {D}iffusion tensor imaging validates
  resting state temporal correlations.
\newblock {\em NeuroImage}, 43:554--561, 2008.

\bibitem{Tass_et_al-98}
P.~Tass, M.G. Rosenblum, J.~Weule, J.~Kurths, A.S. Pikovsky, J.~Volkmann,
  A.~Schnitzler, and H.-J. Freund.
\newblock Detection of $n:m$ phase locking from noisy data: Application to
  magnetoencephalography.
\newblock {\em Physical Review Letters}, 81(15):3291--3294, 1998.

\bibitem{Rodriguez_et_al-99}
E.~Rodriguez, N.~George, J.-P. Lachaux, J.~Martinerie, B.~Renault, and F.~J.
  Varela.
\newblock Perception's shadow: {L}ong distance synchronization of human brain
  activity.
\newblock {\em Nature}, 397(4):430--433, 1999.

\bibitem{Axmacher2008}
N.~Axmacher, D.P. Schmitz, T.~Wagner, Ch.E. Elger, and J.~Fell.
\newblock {Interactions between medial temporal lobe, prefrontal cortex, and
  inferior temporal regions during visual working memory: a combined
  intracranial EEG and functional magnetic resonance imaging study.}
\newblock {\em The Journal of Neuroscience}, 28(29):7304--12, July 2008.

\bibitem{Bartsch26062012}
R.~P. Bartsch, A.~Y. Schumann, J.~W. Kantelhardt, T.~Penzel, and P.~Ch. Ivanov.
\newblock Phase transitions in physiologic coupling.
\newblock {\em PNAS}, 109(26):10181--10186, 2012.

\bibitem{Schreiber-00}
T.~Schreiber.
\newblock Measuring information transfer.
\newblock {\em Phys. Rev. Lett.}, 85(2):461--464, 2000.

\bibitem{PhysRevE.67.055201}
M.~Palu\ifmmode~\check{s}\else \v{s}\fi{} and A.~Stefanovska.
\newblock Direction of coupling from phases of interacting oscillators: An
  information-theoretic approach.
\newblock {\em Phys. Rev. E}, 67:055201, May 2003.

\bibitem{PhysRevE.75.056211}
M.~Palu\ifmmode~\check{s}\else \v{s}\fi{} and M.~Vejmelka.
\newblock Directionality of coupling from bivariate time series: How to avoid
  false causalities and missed connections.
\newblock {\em Phys. Rev. E}, 75:056211, May 2007.

\bibitem{PhysRevLett.99.204101}
S.~Frenzel and B.~Pompe.
\newblock Partial mutual information for coupling analysis of multivariate time
  series.
\newblock {\em Phys. Rev. Lett.}, 99:204101, Nov 2007.

\bibitem{PhysRevE.76.036211}
M.C. Romano, M.~Thiel, J.~Kurths, and C.~Grebogi.
\newblock Estimation of the direction of the coupling by conditional
  probabilities of recurrence.
\newblock {\em Phys. Rev. E}, 76:036211, Sep 2007.

\bibitem{PhysRevLett.100.084101}
A.~Bahraminasab, F.~Ghasemi, A.~Stefanovska, P.~V.~E. McClintock, and H.~Kantz.
\newblock Direction of coupling from phases of interacting oscillators: A
  permutation information approach.
\newblock {\em Phys. Rev. Lett.}, 100:084101, Feb 2008.

\bibitem{PhysRevE.77.026214}
M.~Vejmelka and M.~Palu\ifmmode~\check{s}\else \v{s}\fi{}.
\newblock Inferring the directionality of coupling with conditional mutual
  information.
\newblock {\em Phys. Rev. E}, 77:026214, Feb 2008.

\bibitem{PhysRevLett.100.158101}
M.~Staniek and K.~Lehnertz.
\newblock Symbolic transfer entropy.
\newblock {\em Phys. Rev. Lett.}, 100:158101, Apr 2008.

\bibitem{PhysRevLett.103.238701}
L.~Barnett, A.B. Barrett, and A.K. Seth.
\newblock Granger causality and transfer entropy are equivalent for gaussian
  variables.
\newblock {\em Phys. Rev. Lett.}, 103:238701, Dec 2009.

\bibitem{PhysRevE.83.011919}
M.~Martini, T.A. Kranz, T.~Wagner, and K.~Lehnertz.
\newblock Inferring directional interactions from transient signals with
  symbolic transfer entropy.
\newblock {\em Phys. Rev. E}, 83:011919, Jan 2011.

\bibitem{PhysRevE.83.051122}
B.~Pompe and J.~Runge.
\newblock Momentary information transfer as a coupling measure of time series.
\newblock {\em Phys. Rev. E}, 83:051122, May 2011.

\bibitem{PhysRevE.83.051112}
L.~Faes, G.~Nollo, and A.~Porta.
\newblock Information-based detection of nonlinear granger causality in
  multivariate processes via a nonuniform embedding technique.
\newblock {\em Phys. Rev. E}, 83:051112, May 2011.

\bibitem{Chicharro-Andrzejak-Ledberg-11}
D.~Chicharro, R.~Andrzejak, and A.~Ledberg.
\newblock Inferring and quantifying causality in neuronal networks.
\newblock {\em BMC Neuroscience}, 12(Suppl 1):P192, 2011.

\bibitem{Battaglia-Witt-Wolf-Geisel-12}
D.~Battaglia, A.~Witt, F.~Wolf, and T.~Geisel.
\newblock Dynamic effective connectivity of inter-areal brain circuits.
\newblock {\em PLoS computational biology}, 8:e1002438, 2012.

\bibitem{PhysRevE.86.061121}
J.~Runge, J.~Heitzig, N.~Marwan, and J.~Kurths.
\newblock Quantifying causal coupling strength: A lag-specific measure for
  multivariate time series related to transfer entropy.
\newblock {\em Phys. Rev. E}, 86:061121, Dec 2012.

\bibitem{Kugiumtzis-13}
D.~Kugiumtzis.
\newblock Direct-coupling information measure from nonuniform embedding.
\newblock {\em Phys. Rew. E}, 87:062918, 2013.

\bibitem{Rosenblum-Pikovsky-01}
M.~G. Rosenblum and A.~S. Pikovsky.
\newblock Detecting direction of coupling in interacting oscillators.
\newblock {\em Phys. Rev. E}, 64(10):045202, 2001.

\bibitem{Kralemann_et_al-07}
B.~Kralemann, L.~Cimponeriu, M.~Rosenblum, A.~Pikovsky, and R.~Mrowka.
\newblock Uncovering interaction of coupled oscillators from data.
\newblock {\em Phys. Rev. E}, 76:055201, 2007.

\bibitem{Kralemann_et_al-08}
B.~Kralemann, L.~Cimponeriu, M.~Rosenblum, A.~Pikovsky, and R.~Mrowka.
\newblock Phase dynamics of coupled oscillators reconstructed from data.
\newblock {\em Phys. Rev. E}, 77:066205, 2008.

\bibitem{Cadieu2010}
Ch.F. Cadieu and K.~Koepsell.
\newblock {Phase Coupling Estimation from Multivariate Phase Statistics}.
\newblock {\em Neural Computation}, 22(12):3107--3126, December 2010.

\bibitem{dcm_phase}
W.D. Penny, V.~Litvak, L.~Fuentemilla, E.~Duzel, and K.J. Friston.
\newblock Dynamic {C}ausal {M}odels for phase coupling.
\newblock {\em {J}ournal of {N}euroscience Methods}, 183(1):19--30, 2009.

\bibitem{Kralemann-Pikovsky-Rosenblum-11}
B.~Kralemann, A.~Pikovsky, and M.~Rosenblum.
\newblock Reconstructing phase dynamics of oscillator networks.
\newblock {\em Chaos}, 21:025104, 2011.

\bibitem{PhysRevLett.109.024101}
T.~Stankovski, A.~Duggento, P.~V.~E. McClintock, and A.~Stefanovska.
\newblock Inference of time-evolving coupled dynamical systems in the presence
  of noise.
\newblock {\em Phys. Rev. Lett.}, 109:024101, Jul 2012.

\bibitem{PhysRevE.73.031915}
Z.~Chen, K.~Hu, H.~E. Stanley, V.~Novak, and P.~Ch. Ivanov.
\newblock Cross-correlation of instantaneous phase increments in pressure-flow
  fluctuations: Applications to cerebral autoregulation.
\newblock {\em Phys. Rev. E}, 73:031915, Mar 2006.

\bibitem{Friston-11}
K.~J. Friston.
\newblock Functional and effective connectivity: A review.
\newblock {\em Brain Connectivity}, 1(1):13--24, 2011.

\bibitem{Rubinov-Sporns-10}
M.~Rubinov and O.~Sporns.
\newblock Complex network measures of brain connectivity: {U}ses and
  interpretations.
\newblock {\em NeuroImage}, 52:1059--1069, 2010.

\bibitem{Tass-Rosenblum-Weule-Kurths-Pikovsky-Volkmann-Schnitzler-Freund-98}
P.~Tass, M.~G. Rosenblum, J.~Weule, J.~Kurths, A.~Pikovsky, J.~Volkmann,
  A.~Schnitzler, and H.-J. Freund.
\newblock Detection of n:m phase locking from noisy data: {A}pplication to
  magnetoencephalography.
\newblock {\em Phys. Rev. Lett.}, 81(15):3291--3294, 1998.

\bibitem{Mormann-Lehnertz-David-Elger-00}
F.~Mormann, K.~Lehnertz, P.~David, and C.~E. Elger.
\newblock Mean phase coherence as a measure for phase synchronization and its
  application to the {EEG} of epilepsy patients.
\newblock {\em Physica D}, 144(3-4):358--369, 2000.

\bibitem{Kuramoto-84}
Y.~Kuramoto.
\newblock {\em Chemical Oscillations, Waves and Turbulence}.
\newblock Springer, Berlin, 1984.

\bibitem{Pikovsky-Rosenblum-Kurths-01}
A.~Pikovsky, M.~Rosenblum, and J.~Kurths.
\newblock {\em Synchronization. A Universal Concept in Nonlinear Sciences.}
\newblock Cambridge University Press, Cambridge, 2001.

\bibitem{Kralemann-Pikovsky-Rosenblum-13}
B.~Kralemann, A.~Pikovsky, and M.~Rosenblum.
\newblock Detecting triplet locking by triplet synchronization indices.
\newblock {\em Phys. Rev. E}, 87:052904, 2013.

\bibitem{PhysRevE.71.036207}
D.A. Smirnov and R.G. Andrzejak.
\newblock Detection of weak directional coupling: Phase-dynamics approach
  versus state-space approach.
\newblock {\em Phys. Rev. E}, 71:036207, Mar 2005.

\bibitem{Kreuz200729}
T.~Kreuz, F.~Mormann, R.G. Andrzejak, A.~Kraskov, K.~Lehnertz, and
  P.~Grassberger.
\newblock Measuring synchronization in coupled model systems: A comparison of
  different approaches.
\newblock {\em Physica D: Nonlinear Phenomena}, 225(1):29 -- 42, 2007.

\bibitem{Osterhage_et_al-08}
H.~Osterhage, F.~Mormann, T.~Wagner, and K.~Lehnertz.
\newblock Detecting directional coupling in the human epileptic brain:
  Limitations and potential pitfalls.
\newblock {\em Phys. Rev. E}, 77(1):011914, Jan 2008.

\bibitem{Silfverhuth_et_al-12}
M.J. Silfverhuth, H.~Hintsala, J.~Kortelainen, and T.~Sepp\"anen.
\newblock Experimental comparison of connectivity measures with simulated eeg
  signals.
\newblock {\em Medical and Biological Engineering and Computing},
  50(7):683--688, 2012.

\bibitem{Greenblatt-Pflieger-Ossadtchi-12}
R.E. Greenblatt, M.E. Pflieger, and A.E. Ossadtchi.
\newblock Connectivity measures applied to human brain electrophysiological
  data.
\newblock {\em Journal of Neuroscience Methods}, 207(1):1–16, 2012.

\bibitem{Trentool1}
M.~Lindner, R.~Vicente, V.~Priesemann, and M.~Wibral.
\newblock Trentool: A matlab open source toolbox to analyse information flow in
  time series data with transfer entropy.
\newblock {\em BMC Neuroscience}, 12(1):119, 2011.

\bibitem{Trentool2}
M.~Wibral, R.~Vicente, V.~Priesemann, and M.~Lindner.
\newblock Trentool: an open source toolbox to estimate neural directed
  interactions with transfer entropy.
\newblock {\em BMC Neuroscience}, 12(Suppl 1):P200, 2011.

\bibitem{Seth-10}
A.K. Seth.
\newblock A {MATLAB} toolbox for {G}ranger causal connectivity analysis.
\newblock {\em Journal of Neuroscience Methods}, 186(2):262–273, 2010.

\bibitem{Barnett-Seth-14}
L.~Barnett and A.K. Seth.
\newblock The {MVGC} multivariate {G}ranger causality toolbox: {A} new approach
  to {G}ranger-causal inference.
\newblock {\em Journal of Neuroscience Methods}, 223(2):50--68, 2014.

\bibitem{Shelter_et_al-06}
B.~Schelter, M.~Winterhalder, R.~Dahlhaus, J.~Kurths, and J.~Timmer.
\newblock Partial phase synchronization for multivariate synchronizing systems.
\newblock {\em Phys. Rev. Lett.}, 96(20):208103, May 2006.

\bibitem{Smirnov-Schelter-Winterhalder-Timmer-07}
D.~Smirnov, B.~Schelter, M.~Winterhalder, and J.~Timmer.
\newblock Revealing direction of coupling between neuronal oscillators from
  time series: Phase dynamics modeling versus partial directed coherence.
\newblock {\em Chaos: An Interdisciplinary Journal of Nonlinear Science},
  17:013111, 2007.

\bibitem{Levnajic-13}
Z.~Levnajic.
\newblock Derivative-variable correlation reveals the structure of dynamical
  networks.
\newblock {\em Eur. Phys. J. B}, page 298, 2013.

\bibitem{Levnajic-Pikovsky-14}
Z.~Levnajic and A.~Pikovsky.
\newblock Untangling complex dynamical systems via derivative-variable
  correlations.
\newblock {\em Sci. Rep.}, 4:5030, 2014.

\bibitem{Bashan_et_al-12}
Bashan A., Bartsch R.P., Kantelhardt J.W., Havlin S., and Ivanov~P. Ch.
\newblock Network physiology reveals relations between network topology and
  physiologic function.
\newblock {\em Nature Communications}, 3:702, 2012.

\bibitem{Ivanov-Bartsch-14}
Ivanov P.Ch. and Bartsch R.P.
\newblock Network physiology: {M}apping interactions between networks of
  physiologic networks.
\newblock In D'Agostino G. and Scala A., editors, {\em Networks of Networks:
  the last Frontier of Complexity}, 5394, pages 203--222. Springer
  International Publishing, Switzerland, 2014.

\bibitem{Wagner-Fell-Lehnertz-10}
T.~Wagner, Fell, and K.~Lehnertz.
\newblock The detection of transient directional couplings based on phase
  synchronization.
\newblock {\em New J. of Physics}, 12:053031, 2010.

\bibitem{Levnajic-Pikovsky-11}
Z.~Levnaji\ifmmode~\acute{c}\else \'{c}\fi{} and A.~Pikovsky.
\newblock Network reconstruction from random phase resetting.
\newblock {\em Phys. Rev. Lett.}, 107(3):034101, 2011.

\bibitem{Tenreiro-11}
C.~Tenreiro.
\newblock Fourier series-based direct plug-in bandwidth selectors for kernel
  density estimation.
\newblock {\em Journal of Nonparametric Statistics}, 23(2):533--545, 2011.

\end{thebibliography}

\end{document}